\documentclass[a4paper]{jpconf}
\usepackage{graphicx}
\usepackage{url}

\begin{document}
\title{Low radioactivity techniques for Large TPCs in rare event searches}

\author{Susana Cebrián}

\address{Centro de Astropart\'iculas y F\'isica de Altas Energ\'ias (CAPA), Universidad de Zaragoza, 50009 Zaragoza, Spain
\\ Laboratorio Subterr\'aneo de Canfranc, 22880 Canfranc Estaci\'on, Huesca, Spain}

\ead{scebrian@unizar.es}

\begin{abstract}
The investigation of rare phenomena requires an effective suppression of all the background components entangling the expected signal. This has compelled the development of a wide range of low radioactivity techniques and background mitigation strategies. Some examples of those applied to Large Time Projection Chambers (TPCs) will be discussed here, including the operation of experiments deep underground, the exhaustive control of material radiopurity and the implementation of discrimination techniques.

\end{abstract}

\section{Introduction}

Experiments devoted to investigate rare events, expected to occur with very low probability, like the direct detection of galactic dark matter particles, neutrino interactions or nuclear double beta decays, demand ultra-low background conditions \cite{heusser,formaggio}. Different strategies are being applied for background reduction and mitigation \cite{pdg,snowmassbkg}, depending on several factors like the physics goal of the experiment (considering the type and energy of the expected signal), the detector technology and the particular background components (spanning over a broad range of energies, particle types, and interactions). 

The background events affecting rare event searches have different origins. From cosmic rays, only muons are capable of arriving deep underground, having a flux typically reduced by several orders of magnitude; these residual muons are relevant not only due to their direct interaction but also because of the production of other particles like photons, neutrons or radioactive isotopes. The environmental radiation include neutrons and emissions from radioactivity, either primordial, cosmogenic or even anthropogenic. In addition to $^{40}$K and the natural chains, $^{222}$Rn is particularly relevant due to the concentration in air, material outgassing and plate-out on surfaces of daughter nuclei. Radiogenic neutrons are produced by fission and ($\alpha$,n) reactions in materials, including rock, while cosmogenic neutrons are induced by muons; the energy spectra and fluxes of the two neutron components are different, producing different interactions. The elastic scattering of fast neutrons is very relevant for the direct detection of dark matter, as the induced recoil would be indistinguishable from the expected signal.

After this brief overview of the main background sources affecting the experiments, here several low radioactivity and background mitigation techniques will be described, with a special focus on Large Time Projection Chambers (TPCs) used for rare event searches; a short introduction and a few examples of application will be presented for each technique. In order to reduce the flux of background-creating radiation, the use of passive shieldings, the operation underground and a thorough control of material radiopurity are mandatory, as described in Sections~\ref{secund} and \ref{secrad}. Once background is unavoidable, an active event tagging to identify signals can be made by using active shieldings or developing background rejection techniques, as shown in Section~\ref{secdis}. In the whole process of understanding and mitigating backgrounds, Monte Carlo simulation is a valuable tool; several available codes, elaborated with different purposes, will be presented in Section~\ref{secsim}.

\section{Undeground facilities and shieldings}
\label{secund}

\begin{figure}
\begin{center}
\includegraphics[width=\textwidth]{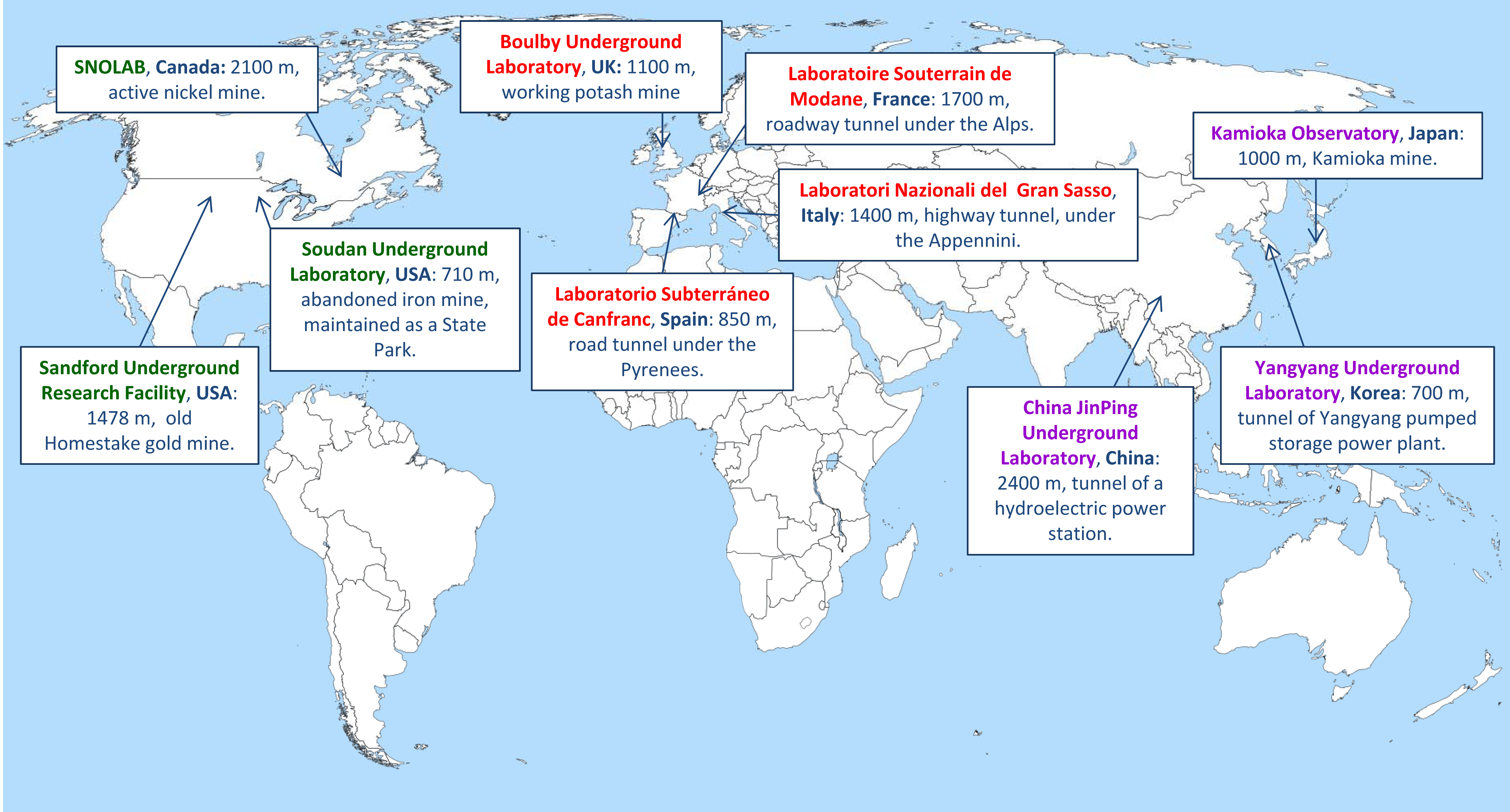}
\end{center}
\caption{\label{map} Location, depth and other features of the main underground laboratories worldwide in operation over the last years.}
\end{figure}

At deep underground laboratories, with a rock overburden typically larger than 1000 meters of water equivalent (m.w.e.), the flux of cosmic muons is reduced by several orders of magnitude and the rest of secondary cosmic radiation arriving to the Earth's surface is totally suppressed. These research infrastructures provide not only underground space to the experiments but also services (for chemistry, mechanics, electronics or cryogenics) to users and specialized facilities intended to reduce and control the radioactive background, like clean rooms, radon abatement systems, or different types of radioassay facilities \cite{ufianni,uftaup}. Figure \ref{map} shows the geographic location and main features of ten underground facilities in operation over the last years, located in Asia, Europe and North America. Although not shown in the map, other underground laboratories are WIPP in US, Rustrel in France, Baksan in Russia, Callio Lab in Finland, and LABchico in Mexico; in addition, Yemilab in Korea and Stawell Laboratory in Australia are just starting operation, the latter being the first underground laboratory in the southern hemisphere. There are plans to build other underground facilities in India, South Africa and the ANDES laboratory between Chile and Argentina. These facilities are becoming multidisciplinary research infrastructures, expanding rapidly in total area and space; an overview of present and future underground facilities and infrastructure is presented at \cite{ufsnowmass}.

Getting rid of cosmic radiation is not enough, as environmental radiation from neutrons and radioactivity is also present deep underground. The installation of passive shieldings around the detectors is necessary, combining high-Z materials (like lead or copper) for gamma background and low-A materials (like water or polyethylene) to moderate neutrons.
Active shielding based on different approaches is necessary in many cases. Segmented detectors allow the rejection of coincident events, which must be produced by background. The definition of a fiducial volume is possible in detector systems having spatial resolution in order to suppress background events produced on the surfaces surrounding the detector volume. Veto systems for external gamma radiation, neutrons and muons can be implemented using plastic scintillators, liquid scintillators or water Cerenkov detectors placed around the principal detector.

As an example of sophisticated active shielding systems, the case of neutron vetos for large dark matter detectors using TPCs can be mentioned. In the DarkSide-20k experiment, using liquid Argon (LAr), the TPC walls will be made of Gd-loaded acrylic, being placed between two LAr buffers read by SiPMs \cite{ds}; in this way, neutrons thermalized and captured on  Gd will be tagged by detecting the emitted gamma rays. In the XENONnT and LUX-ZEPLIN experiments, using liquid Xe, the neutron tagging capability will be obtained thanks to a tank of Gd-loaded water or liquid scintillator, respectively, surrounding the Xe TPC.

\section{Material radiopurity} 
\label{secrad}

Rare event experiments typically demand materials with activity of primordial isotopes ($^{238}$U, $^{232}$Th, $^{40}$K) at or below mBq/kg, which is 3-4 orders of magnitude lower that the usual activity found in the Earth's crust. This requirement imposes a careful selection of materials and components for the design of all the parts of the set-up, from the results of radioassays. The measured activities are used in addition as inputs for the development of background models based on Monte Carlo simulations to quantify the sensitivity of experiments. It is recognized that the testing and certification of materials (considering different providers and even production batches) involve precision measurements consuming significant amounts of time and money.

Several techniques are routinely applied to quantify ultra-low levels of activity in selected samples. Table \ref{comp} shows a summary of pros and cons of the main techniques. Mass Spectrometry (Glow Discharge Mass Spectrometry, GDMS) or Inductively Coupled Plasma Mass Spectrometry, ICPMS) provide concentrations of U, Th and K, which allow to derive the activity of the mothers of the natural chains \cite{ms}; GDMS is suitable only for metals. Neutron Activation Analysis (NAA) is also used \cite{naa}, although the access to nuclear reactors is not simple and a careful preparation of the irradiation is necessary. Measurements based on gamma-ray spectroscopy with ultra-low background HPGe detectors are made at almost all the underground laboratories (see for instance \cite{gamma1,gamma2}); being a non-destructive technique, the components to be actually used in the experiment can be analyzed. Alpha spectroscopy, combined in some cases with radiochemical methods, helps to quantify the activity of isotopes in the lowest part of the $^{238}$U chain, where gamma spectroscopy is not sensitive. Not only the activity in the bulk can be relevant, but also that on material surfaces as well as the effect of Radon emanations; therefore, specific methods to quantify this are in development too.

\begin{center}
\begin{table}[h]
\caption{Comparison of advantages and disadvantages of the main techniques used to quantify ultra-low activity levels in materials used in rare event experiments.} \label{comp}
\centering
\begin{tabular}{l|l|l}
\br
Technique & Advantages & Disadvantages \\
\mr
Mass spectrometry & Fast & Destructive\\
& Small samples required & Not sensitive to chain disequilibrium\\ \hline
NAA & Excellent sensitivity & Samples activated \\
 & & Irradiation not straightforward \\ \hline
Ge $\gamma$ spectroscopy & Non-destructive & Large samples required \\
& Sensitive to relevant isotopes & Long measurements required 
\\
\br
\end{tabular}
\end{table}
\end{center}

Many experiments have shared the results from their material screening programs (see for example those related to TPCs \cite{exo,next,xenon,lz,trexdm}). Attempts of creating single public repositories of radiopurity measurements have been made and now a complete database is available at \url{https://www.radiopurity.org } \cite{database}.

A selection of available radiopure materials may not be enough and in some cases special procedures must be applied to actively reduce the material activity. Some purification methods for bulk activity are distillation, zone melting for crystals or copper electroformation \cite{brodzinski}. The effect of Radon daughters plate-out on surfaces forces to apply different cleaning protocols to remove a thin layer of material, from clean machining, acid etching or electropolishing to more sophisticated techniques; they have been used on metals like copper or stainless steel and on plastics like teflon, reducing very effectively the surface activity \cite{hoppe,stein,bruenner}. Cleaning protocols must be developed paying special attention to the radiopurity of chemicals and other components (gloves, wipes, \dots) needed; even the control of particulate fallout and assembly in clean rooms may be required \cite{vacri}.

Production of long-lived radioactive isotopes in materials due to exposure to cosmic rays (mainly by spallation) during production, transport and storage of components, can become an hazard for ultra-low background experiments \cite{cosmog1,cosmog2}. Avoiding flights, storing materials underground to limit surface residency time and even using shields against cosmic rays mitigate the cosmogenic activation; but as these requirements typically complicate the experiment preparation, it is advisable to quantify the real danger of the material activation. To estimate the induced activity, knowing the exposure history to cosmic rays, the production rates of relevant isotopes in the targets must be known, either from scarce experimental data (from irradiation or experiments with controlled exposure) or from calculations (using production cross sections and the cosmic ray spectrum). For instance, cosmogenic activation in Argon has been recently studied \cite{saldanha,dsact}.

The stringent requirements of material radiopurity in rare event experiments have made necessary the development of custom-made detectors to quantify very low levels of activity in particular situations. The following are some examples of such special detectors.

\begin{itemize}
\item The BiPo detector \cite{bipo} was built by the SuperNEMO collaboration and operated for several years in the Canfranc Underground Laboratory, being sensitive to activities at the level of $\mu$Bq/kg of isotopes at the lower parts of $^{238}$U and $^{232}$Th chains thanks to the identification of the so-called BiPo $\alpha$-$\beta$ coincidences. The very thin samples were placed in a sandwich of plastic scintillator detectors. Not only the foils with double beta emitters to be used in the SuperNEMO demonstrator were assayed, but also other thin samples like Micromegas readouts; for them, upper limits below 0.1~$\mu$Bq/cm$^2$ were derived \cite{trexdm}.
\item The AlphaCAMM detector (Alpha CAMera Micromegas) \cite{alphacamm} is being built at the University of Zaragoza; a gaseous TPC read with a segmented Micromegas at the anode is intended to measure $^{210}$Pb surface contamination of flat samples placed at the cathode, down to 100~nBq/cm$^2$. Micromegas readouts can provide topological information to reconstruct origin and end of $\alpha$ tracks from $^{210}$Po, allowing a better identification of decays actually produced at the sample. After the proof-of-concept with a non-radiopure prototype, a radiopure detector is being commissioned.
\item The procurement of low-radioactivity underground Ar (UAr) for the DarkSide-20k experiment has required the development of special facilities: Urania, in Colorado (US), for the extraction of UAr from CO$_2$ wells; Aria, in Sardinia (Italy), for the purification in a cryogenic distillation column \cite{aria}; and the DArT detector, in Canfranc (Spain), for the quantification of $^{39}$Ar activity. The DArT detector (Depleted Argon Target) \cite{dart} is intended to measure $<1$~mBq/kg of $^{39}$Ar with statistical accuracy better than 10\% in one week of counting time. It consists of a small chamber read with SiPMs placed inside a $\sim$1~t atmospheric Ar buffer acting as shield and veto.
\end{itemize}

\section{Discrimination techniques} 
\label{secdis}

Once the background sources become unavoidable, it is necessary to develop, when possible, mechanisms to discriminate on an event-by-event basis the expected signal from backgrounds; they are indistinguishable if only their energy deposition is measured, but event topologies may differ significantly in time and space. More and more sophisticated techniques are being applied, based on many different approaches depending on the detection technology.

Pulse Shape Discrimination (PSD) is possible in some detectors to identify and reject PMT noise o for Particle Identification (PID) to separate neutrons, alpha particles or beta/gamma emissions. The distribution of different parameters evaluated from the pulses for the different event populations allow to establish cuts for the discrimination; in this process, it is essential to evaluate carefully the cut efficiency. Multiparametric cuts can be defined and even combined applying machine learning techniques. As an example, Argon exhibits a different scintillation pulse shape for electronic and nuclear recoils (ER/NR), which allows a very efficient PSD in dark matter detectors; single phase detectors, where only scintillation is measured, have both PSD and position reconstruction. This is the approach of the DEAP experiment in SNOLAB, having derived results with an astonishing ER leakage probability of 10$^{-9}$ \cite{deap}.

For the direct detection of dark matter particles, the simultaneous measurements of heat and light/ionization signals in the so-called hybrid detectors provides a very effective discrimination between ER and NR, as the relative yield of heat and charge/light signals is different. This has been successfully applied in semiconductor and scintillating bolometers. Cryogenic scintillators, such as Xe and Ar, used in dual-phase TPCs offer a strong ER background mitigation through the detection of a primary scintillation signal in liquid (S1) and a secondary signal in gas from the drift and extraction of ionization electrons (S2). Electron and nuclear recoils have different ionization and quenching effects and this allows a strong discrimination. 

For double beta decay experiments, the analysis of the topology of events provides very useful information. If position reconstruction is possible for the energy deposits in the detector, mono-site events, expected for the signal, can be discriminated from multi-site events, assigned to backgrounds. Moreover, in detector with tracking capabilities an efficient background rejection can be made looking for the tracks from a common vertex of the two electrons emitted in the double beta decay. This has been demonstrated by the NEMO3 experiment, using a combination of plastic scintillators and Geiger cells. The NEXT project is focused on the operation of a High Pressure Xe-Electroluminiscence gas TPC with separate energy and tracking readout, applying different event reconstruction and selection methods \cite{nextpau}.

\section{Simulation and codes} 
\label{secsim}

Monte Carlo simulations of the interaction of background radiation in matter, reproducing particle transport and detector response, are extremely useful in several ways: for the understanding of measured data and the construction of background models, for the assessment of the effect of background reduction strategies (shieldings, vetoes, discrimination, \dots) and for the evaluation of the expected counting rates and the sensitivity to the investigated physical processes of future experiments.

Thera are general-purpose packages extensively used in many different contexts like GEANT4 (GEometry ANd Transport) \cite{geant4}, FLUKA \cite{fluka}, or MCNPX \cite{mcnp}. They allow to define the primary particles in the simulation, the physical processes to be considered, the experimental set-up and the outputs to be obtained. Stopping and Range of Ions in Matter (SRIM) is a Monte Carlo package that calculates interactions between ions and matter \cite{srim}. Specific Geant4 environments have been created and validated against data for particular technologies, like MaGe, G4DS or ImpCRESST, and the detector response has been integrated by specific codes/modules: NEST (Noble Element Simulation Technique) for the simulation of the scintillation, ionization, and electroluminescence processes in noble elements \cite{nest}, G4CMP (Condensed Matter Physics with Geant4) for solid-state detectors \cite{g4cmp} and REST-for-Physics (Rare Event Searches Toolkit for Physics) using gaseous TPCs for data analysis and characterization \cite{rest}.


Specific tools, like codes, libraries or databases, focused on particular backgrounds are also available and in development. A few examples are presented in the following.

\begin{itemize}
    \item EXPACS (EXcel-based Program for calculating Atmospheric Cosmic-ray Spectrum) \cite{expacs} allows to calculate terrestrial cosmic ray fluxes of neutrons, protons, light ions, muons, electrons, positrons, and photons nearly anytime and anywhere in the Earth's atmosphere. The CRY (“Cosmic-ray Shower Library”) generator \cite{cry} gives also energies, positions, and directions of different cosmic particles.
    \item To obtain muon intensity and energy and angular distributions of muons there are several codes like MUSUN, MUSIC \cite{musicmusun} and MUTE (MU inTensity codE) \cite{mute}.
    \item Quantifying the yields of radiogenic neutrons, produced by spontaneous fission of $^{238}$U and ($\alpha$,n) reactions, is very important, in particular, for dark matter experiments. This allows to set requirements on the activity of materials. Several codes are available, following different approaches for the transport of the $\alpha$ particles in the materials and for the estimate of the reaction cross sections. In SOURCES4A(C) \cite{sources}, the cross sections are obtained from the EMPIRE2.19 code, while in the USD calculator \cite{mei09} the TALYS code is used. TALYS is also used, combined with SRIM for the stopping power of $\alpha$ particles at NeuCBOT (Neutron Calculator Based On TALYS) \cite{neucbot}. In SaG4n \cite{mendoza}, Geant4 is considered for $\alpha$ transport together with data libraries (JENDL, TENDL) to model ($\alpha$,n) reactions.
    \item The estimate of cosmogenically induced activity in materials requires the knowledge the production cross sections in the targets. Experimental data from beam experiments, typically scarce for neutrons, can be found at the EXFOR (Experimental Nuclear Reaction Data) database \cite{exfor}. Cross sections can be deduced from semiempirical formulae, like the Silberberg\&Tsao equations \cite{tsao1,tsao2,tsao3} for targets with mass number $A\geq3$, products with $A\geq6$ and energies larger than $\sim$100 MeV; these formulae have been implemented in several codes like COSMO \cite{cosmo}, YIELDX \cite{tsao3} and ACTIVIA \cite{activia}. Alternatively, the cross section can be derived from Monte Carlo simulation, including the formation and decay of compound nuclei and de-excitation processes like fission, fragmentation, spallation, or break-up. Many different models and codes have been developed in very different contexts (studies of comic rays and astrophysics, transmutation of nuclear waste or production of medical radioisotopes, for instance). Several libraries compile results obtained with different codes: TENDL (TALYS-based Evaluated Nuclear Data Library) \cite{tendl} includes data for both proton and neutrons up to 200~MeV as projectiles, derived from TALYS; JENDL (Japanese Evaluated Nuclear Data Library) \cite{jendl} is based on the GNASH code offering cross sections for neutrons and protons up to 3~GeV; HEAD-2009 (High Energy Activation Data) library~\cite{head2009} includes results for protons from 150~MeV to 1~GeV obtained using a selection of models and codes. As discrepancies in cross sections from different approaches are usually not negligible, it is a good practice to collect information from different sources of data, taken into account both calculations and measurements if possible.    
\end{itemize}

\section{Summary} 

Rare event searches demand ultra-low background conditions, which can be achieved by mitigating all known background sources: cosmic muons, radiogenic and cosmogenic neutrons, and radioactivity. Deep underground laboratories placed all over the world provide not only shelter for cosmic rays but different, specialized support facilities, for example, for material radioassay. Passive and more sophisticated active shieldings are a must to reduce or discriminate backgrounds, together with specific rejection techniques, based on different approaches, which are in continuous development. Analysis methods based on machine learning techniques are very effective to discriminate background/noise events and Monte Carlo simulations based on different codes are valuable tools to study backgrounds. As the background level of experiments has been getting lower and lower over time, experimental techniques to understand, measure and suppress detector backgrounds have been improved and nowadays it is really a frontier aspect for rare event searches \cite{pdg,snowmassbkg}.

\end{document}